\documentclass[fleqn,usenatbib]{mnras}

\usepackage{newtxtext,newtxmath}

\usepackage[T1]{fontenc}

\DeclareRobustCommand{\VAN}[3]{#2}
\let\VANthebibliography\thebibliography
\def\thebibliography{\DeclareRobustCommand{\VAN}[3]{##3}\VANthebibliography}

\usepackage{graphicx}	
\usepackage{amsmath}	

\title[R\'enyi entropy]{Stability analysis of planetary systems via second-order R\'enyi entropy}

\author[T. Kov\'acs et al.]{
Tam\'as Kov\'acs,$^{1,2}$\thanks{E-mail: tamas.kovacs@ttk.elte.hu(TK)}
M\'at\'e Pszota\,$^{1,2}$ 
Emese K\H{o}v\'ari\,$^{2,3,4}$ 
Emese Forg\'acs-Dajka\,$^{2,3,4,6}$ 
and Zsolt S\'andor\,$^{3,4,5}$
\\
$^{1}$ELKH-ELTE Extragalactic Astrophysics Research Group, Institute of Physics, E\"otv\"os Lor\'and University, Budapest, Hungary\\
$^{2}$Wigner Research Centre for Physics, P.O. Box 49, Budapest H-1525, Hungary\\
$^{3}$Department of Astronomy, Institute of Geography and Earth Sciences, E\"otv\"os Lor\'and University, H-1117 Budapest, P\'azm\'any P\'eter s\'et\'any 1/A, Hungary\\
${^4}$ Centre for Astrophysics and Space Science, E\"otv\"os Lor\'and University, H-1117 Budapest, P\'azm\'any P\'eter s\'et\'any 1/A, Hungary\\
${^5}$ Konkoly Observatory, Research Centre for Astronomy and Earth Sciences, Budapest, Hungary\\
${^6}$ ELKH-SZTE Stellar Astrophysics Research Group, H-6500 Baja, Szegedi út, Kt. 766, Hungary
}

\date{Accepted XXX. Received YYY; in original form ZZZ}

\pubyear{2015}

\begin{document}
\label{firstpage}
\pagerange{\pageref{firstpage}--\pageref{lastpage}}
\maketitle

\begin{abstract}
The long-term dynamical evolution is a crucial point in recent planetary research. Although the amount of observational data is continuously growing and the precision allows us to obtain accurate planetary orbits, the canonical stability analysis still requires N-body simulations and phase space trajectory investigations. We propose a method for stability analysis of planetary motion based on the generalized R\'enyi entropy obtained from a scalar measurement. The radial velocity data of the central body in the gravitational three-body problem is used as the basis of a phase space reconstruction procedure. Then, Poincar\'e's recurrence theorem contributes to finding a natural partitioning in the reconstructed phase space to obtain the R\'enyi entropy. It turns out that the entropy-based  stability analysis is in good agreement with other chaos detection methods, and it requires only a few tens of thousands of orbital period integration time.  
\end{abstract}



\section{Introduction}

Long-term orbital stability of the gravitational few body problem is a fundamental question in contemporary planetary research. The dynamical evolution of the solar system, as well as newly discovered exoplanetary worlds, serves as fundamental knowledge about the formation, architecture, and possible future configurations of these systems. 

Different kinds of stability criteria, such as Lyapunov, Hill, Lagrangian have been considered while studying the dynamical evolution of planetary systems. Also, various phenomena of chaotic behaviour (strong and weak chaos, Chirikov \citep{Chirikov1979} and Nekhoroshev \citep{Nekhoroshev1977} regimes) emerged from the fundamental theoretical works dated back to the mid-1900s. 
Quantifying the stability of a dynamical system, including planetary motion, can be done by using an analytic approach, particularly of resonant dynamics \citep{Wisdom1980,Laskar1989,Duncan1993,Celetti2006,Robutel2006,Batygin2008} or calculating a variety of numerical chaos indicators (MEGNO \citep{Cincotta2000}, RLI \citep{Sandor2004}, SALI/GALI \citep{Skokos2004,Skokos2007}) most of them based on trajectory divergence in phase space.  
Basically, the instability of a system is manifested in the diffusion of action-type variables. More concretely, the time evolution of the variance in actions carries relevant information about dynamics. Recently, alternate methods have been proposed to describe system stability \citep{Cachucho2010,Marti2016} according to the statistical description of action variables. 

Although, the above-mentioned methods are efficient and provide a reliable representation of long-term dynamics, the need for computationally cheaper numerical schemes, i.e. shorter integration, is formulated. New studies reveal that information-theoretic concepts such as Shannon entropy can be used to characterise not even the stability but the average lifetime of a particular planetary system with unbounded phase space \citep{Cincotta2018,Beauge2019,Cincotta2021,Kovari2022}. The authors of these studies claim that in the case of entropy calculations a very short integration is enough to tackle a problem of orbital stability. Moreover, their results are promising to estimate the system's lifetime deduced from the diffusion coefficient that is also calculated from the entropy of a trajectory. 

Our present work fits the above efforts. Namely, we want to analyze the stability of planetary systems via an entropic quantity, the R\'enyi entropy of second order (see Section \ref{sec_ent} for more details). The method presented in this paper, however, does not require the knowledge of the whole phase space trajectory nor the equations of motion. In other words, the stability of the system will be gained from measurements (practically scalar time series) rather than numerical integration. 
This viewpoint supports the fact that in exoplanetary observations we have limited knowledge about the full dynamics. That is, one can measure the starlight and its systematic variations (radial velocity, transit times, astrometric positions) due to the surrounding planets to figure out the system parameters and dynamical properties.  

The theoretical background that allows us to obtain R\'enyi entropy from an experimental data set is Poincar\'e's recurrence theorem along with the visualisation tool of recurrence plots \citep{Eckmann1987} and their statistical quantification \citep{Webber2005}, see Section~\ref{sec_methods}. In fact, this technique is a powerful tool in nonlinear time series analysis and the exploration of underlying dynamics. We shall note at this point that recurrence quantification analysis has been used in various astronomical problems for more than a decade, e.g. long-term variation of sunspots  \citep{Zolotova2009}, dimming and brightening events of variable star Betelgeuse \citep{George2008}, temporal  light curve changes of black hole GRS1915+105 \citep{Harikrishnan2011}. \cite{Tsiganis2000} showed that the second-order correlation sum and the associated correlation dimension are suitable to distinguish filling chaotic and sticky orbits with statistical significance. \cite{Floss2018} studied the dimensionality of the recently discovered planetary system TRAPPIST-1. Moreover, \cite{Asghari2004} used R\'enyi entropy to construct semimajor-axis -- eccentricity stability maps of several exoplanetary systems (Gl 777 A, HD 72659, Gl 614, 47 Uma, HD4208). They demonstrate that the recurrence based technique gives the same result as the widely used maximum eccentricity method. However, their calculations involve the entire phase space trajectory. Our primary goal is to show that R\'enyi entropy is suitable to identify different dynamical regimes for the same length of integration (MEGNO - full phase space trajectory) and data set (scalar radial velocity measurement of the sun).

This paper is organized as follows. In the next section, we give the details of the dynamical system investigated through our study. Section~\ref{sec_methods} is devoted to the methods how to reconstruct a trajectory from scalar time series, build up recurrence plots and calculate R\'enyi entropy. Our results are presented in Section~\ref{sec_results} followed by a discussion. 

\section{The model system}

\begin{figure}
\begin{center}
\includegraphics[width=\columnwidth]{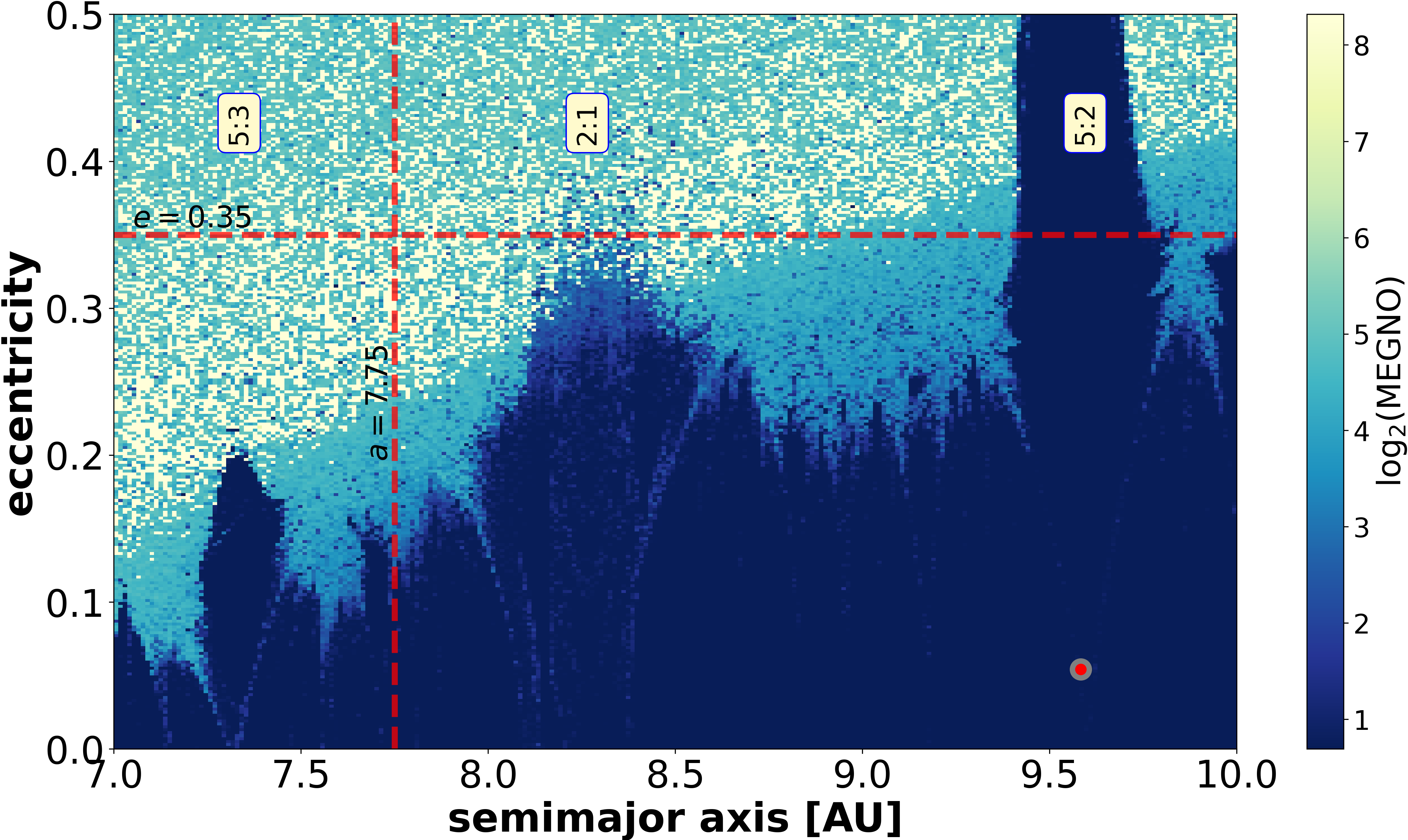}
\end{center}
\caption{The ($a,e$) phase portrait of the Sun-Jupiter-Saturn system. Each point represents an initial condition and the associated stability measure according to the colour bar on the right. Larger values ($\mathrm{MEGNO}>2$) indicate chaotic motion. The range of MEGNO has been cropped at 10. The horizontal and vertical dashed lines, $e=0.35$ and $a=7.75,$ respectively, guide the eye at further analysis, see the text in Section~\ref{sec_results}. The red circle illustrates Saturn's real position. The stability map as well as the sun's radial velocity data is computed over 1000 Jupiter periods (i.e. $\sim$ 1.2e4 years.).}\label{fig:1}
\end{figure}

We use the Sun-Jupiter-Saturn (SJS) general three-body problem to have a benchmark in stability analysis. Figure~\ref{fig:1} depicts the familiar semimajor-axis--eccentricity portrait of system stability. It has been generated as follows. We implemented the barycentric coordinates (J2000) of the sun and the planets. Then integrated the system by the WHFAST scheme of the \texttt{Rebound} package \citep{Rein2015} starting from 250x250=62500 different initial conditions of Saturn resulting in a fine grid of $a_{\mathrm{Saturn}}=[7;10],$ $e_{\mathrm{Saturn}}=[0;0.5].$\footnote{For simplicity, we omit the indication that orbital elements refer to Saturn in the rest of the paper.} After 10000 Jupiter orbital periods each pixel of the ($a,e$) map is colour coded accordingly to the chaos indicator MEGNO obtained from the numerical integration, too. The lower (dark) part ($e\lesssim 0.2$) of the plot denotes stable configurations (MEGNO=2). While larger eccentricities due to stronger interaction between the planets lead to chaotic motion. One can also identify mean-motion commensurabilities (MMCs) of Jupiter and Saturn (5:3, 2:1, 5:2 from left to right) that might protect the system against irregular behaviour even for more elongated orbits.            

Once having the access to the full phase space trajectory in the SJS system, we can mimic real astronomical observations. For instance, keeping only the x component of the Sun's velocity, it can be considered as the radial velocity viewed from a distant position outside the solar system exactly along the x-axis. Similarly, from numerical integration, one can acquire quantities like transit timing variation of either planet or celestial position of the central star. These synthetic data as scalar time series will then be the base of our calculations. Particularly, in this study, we use the Sun's velocity component as the only information source about the SJS dynamical system and reconstruct the state vector in order to be able to determine the R\'enyi entropy as a stability index.

\section{Methods}
\label{sec_methods}

An important element in the theory of dynamical systems is Poincar\'e's recurrence theorem. The essence of this is that in the phase space of any conservative system, the trajectories return to a predetermined radius environment of a previous state after a certain time (Poincar\'e return time) \citep{Altmann2005}. This previous state can obviously be the initial condition from which the motion started or an arbitrarily chosen subset of the phase space. The return and the return time, of course, depend on the initial condition of the given trajectory and the desired accuracy. Statistical analysis of return times has now become essential for understanding the basic dynamic properties (phase-space diffusion, ergodicity, irregularity) of low-dimensional systems.

\subsection{Phase space reconstruction}

The dynamical analysis of a particular system requires the evolution of phase space trajectories. Since in an experimental setting the observer records the signals in the time domain, which means not all relevant components of the state vector are known, a reconstruction of multidimensional information in an artificial phase space is needed. This is possible when the following assumption holds \citep{Semmlow}: the variable corresponding to the observable affects the other state variables, i.e. the variables governing the system's dynamics are coupled. In other words, all hidden variables in the system make some
contribution to the measured signal. In this case, a  
recovery of the phase space trajectory can be done by using an embedding theorem \citep{Takens1981,Mane1981,Packard1980}. In what follows, the method of time delay reconstruction is interpreted. 

Time series $x(t_i)$ is a sequence of $i=1,\dots,\;n$ scalar measurements of some
quantity ($x$) depending on the state of the system ($\mathbf{s}$) taken at discrete times ($\Delta t$), that is $x(t_i) = x(\mathbf{s}(i\Delta t)).$ The reconstructed $m$ dimensional vector reads \citep{Kantz2003} then
\begin{equation}
x(t_i) \rightarrow \mathbf{x}_N = \{x(t_i-(m-1)\tau), x(t_i-(m-2)\tau),\dots, x(t_i-\tau), x(t_i))\},
\label{eq:tde}
\end{equation}
where $i=1\dots n$ is the length of the original signal, $m$ is the dimension into which the reconstructed vector is embedded, the delay, $\tau,$ is the time difference between adjacent components of $\mathbf{x}.$ After the delay reconstruction, the length of $\mathbf{x}$ becomes $N=n-(m-1)\tau.$ That is, the components of the reconstructed vector are segments of the original 1D signal delayed by $\tau.$ Thus, $\mathbf{x}$ is an $m\times [n-(m-1)]\tau$ matrix. The optimal delay parameters have been calculated by the \texttt{TISEAN}\footnote{\url{https://www.pks.mpg.de/~tisean/Tisean_3.0.1/}} software package \citep{Hegger1999} through the whole analysis.

The method of time delay embedding allows one to transform an observed scalar time series into a higher dimensional representation in the embedding space. The advantage of this technique is that it provides a
one-to-one image of the original (multi-dimensional) state vector $\mathbf{s}$. This procedure is essential in order to scrutinize Poincar\'e recurrences in high dimensional (reconstructed) phase space. It should, however, be noted that the reconstructed trajectory is not identical to the one we would get from numerical integration, i.e. when all the components are known. It might differ in its shape but preserves the dynamical properties such as topology and Lyapunov exponents. 


\subsection{Recurrence plots}
\label{sec_rp}

Once the phase space trajectory is available, either the original (e.g. from numerical simulation) or the reconstructed one (from experimental measurement),  the return times can be evaluated. The graphical representation of these events, called recurrence plot (RP), has been proposed by \cite{Eckmann1987}. 

An RP can be represented by a two-dimensional binary matrix 
\begin{equation}
\mathbf{R}(\epsilon)=\Theta(\epsilon-||\mathbf{x}_i-\mathbf{x}_j||),\quad
i,j=1\dots N,
\label{eq:RP}
\end{equation}
where $N$ is the length of the (reconstructed) phase space trajectory, $\Theta(\cdot)$ is the Heaviside step function, $\epsilon$ a tolerance parameter, and $||\cdot||$ is a norm. The embedded delay vectors obtained from measured points are $\mathbf{x}_i$ and $\mathbf{x}_j$ at different time instants, $t_i$ and $t_j,$ respectively. If a trajectory at $t_j$ returns to an $\epsilon$ neighbourhood of a point where it was at $t_i$ ($t_j>t_i$) then the corresponding matrix element is 1, i.e. recurrence occurs, otherwise 0.

\begin{figure}
\begin{center}
\includegraphics[width=\columnwidth]{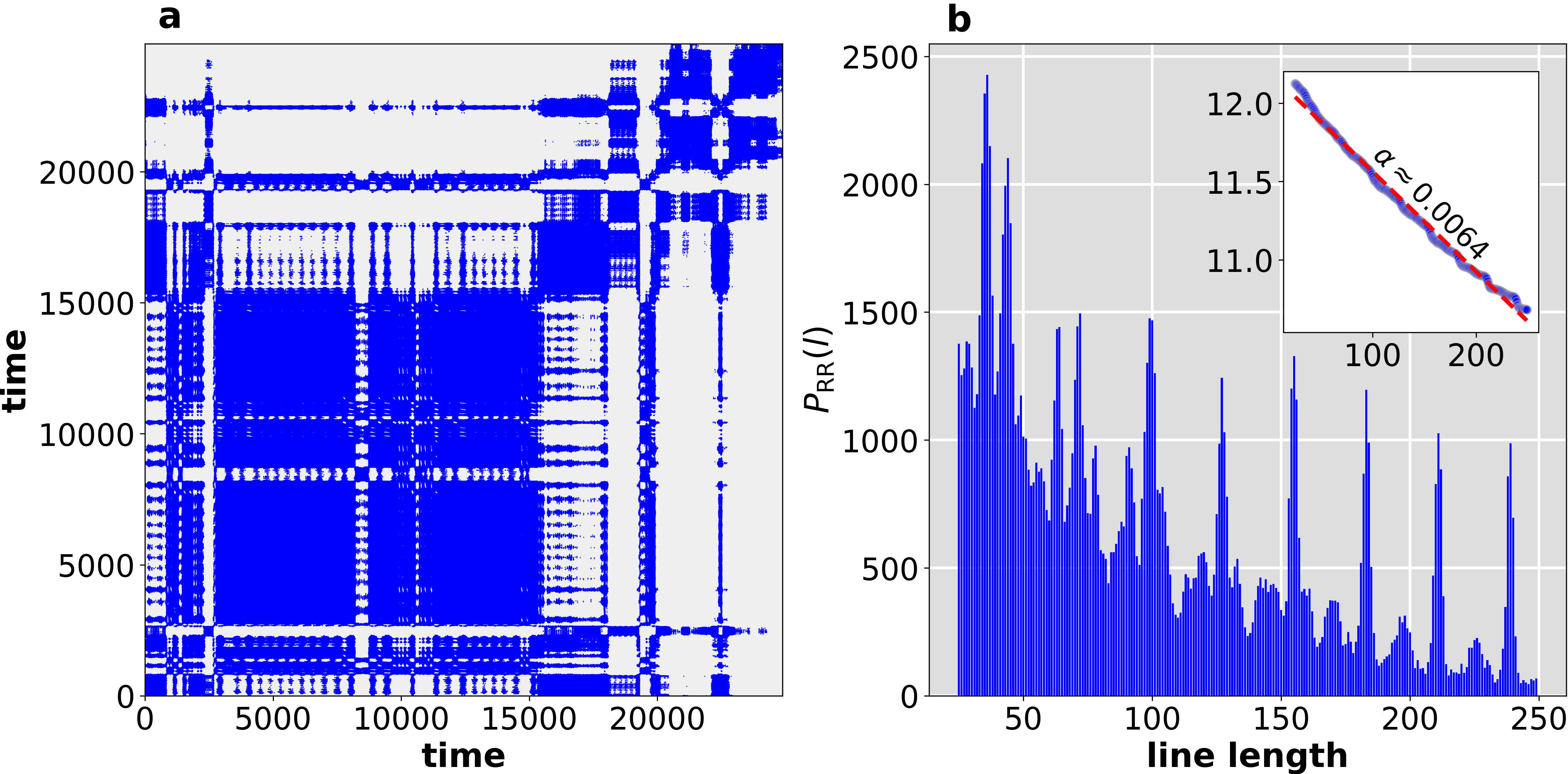}
\end{center}
\caption{(a) Recurrence plot from the Sun's radial velocity. The blue dots display the time instants when the trajectory returns close to its past state in reconstructed phase space. The integration time is 10000 Jupiter periods, N=25000 data points, RR=0.05. The motion represented by this RP has a chaotic nature as indicated by the position of its initial conditions and MEGNO value in Fig.~\ref{fig:1}. (b) $P_{\mathrm{RR}}(l)$ is the number of diagonals of length $l$ in RP (up to $l=250$). The straight line with slope $\alpha$ (inset) is an estimator of second order R\'enyi entropy (see Section~\ref{sec_ent}). Note that the inset shows the semi-logarithmic plot of the \textit{cumulative} $P_{\mathrm{RR}}(l)$ vs. $l$, i.e. number of diagonals of at least length $l.$}\label{fig:2}
\end{figure}

The matrix $\mathbf{R}$ is symmetric by definition. Plotting the elements of the binary matrix with different colours, one can obtain the RP. Figure~\ref{fig:2}a depicts a recurrence plot corresponding to the initial conditions $(a,e)=(7.75,0.35)$ (the intersection of two red dashed lines) in Fig.~\ref{fig:1}. In Fig.~\ref{fig:2}a, the RP corresponds to the reconstructed
trajectory from the Sun's RV data. Basically, RPs have different patterns associated with different kinds of the time evolution of trajectories (periodic,
quasi-periodic, chaotic) \citep{Zou2007,Ngamga2012}. At the very basic level, dots typically appear as diagonal line segments on the RP. Periodic signals yield non-interrupted lines, while chaotic dynamics results in a pattern of diagonals much shorter than for periodic cycles. For a profound review about different structures of RPs see Refs.~\citet{Marwan2004,Marwan2007}.




Since the RP is an intuitive visualization tool to have some impression about the underlying dynamics, it is worth quantifying the structures in it using recurrence quantification analysis (RQA) \cite{Zbilut1992}. The texture of an RP can be classified into single dots, diagonal lines, and horizontal and vertical lines. In what follows,  some RQA measures are demonstrated. 

Recurrence rate ($RR$) is the measure of the density of recurrence points in the RP, i.e. the non-zero entries of $\mathbf{R}$
\begin{equation}
  RR=\frac{1}{N^2}\sum_{i,j=1}^{N}R_{i,j}.
  \label{eq:rr}
\end{equation}

Diagonal line segments of length $l$ characterize those parts of trajectories that remain close to each other at different times for $l$ time steps. Consequently, the larger the $l,$ the longer the time as long as the trajectory copies itself. Thus, defining an average diagonal line length can be used to estimate  the mean prediction time.
\begin{equation}
L=\frac{\sum_{l=l_{\text{min}}}^{N}lP(l)}{\sum_{l=l_{\text{min}}}^{N}P(l)},
\end{equation}
$P(l)$ is the histogram of the line segments of a given length $l$ (Figure~\ref{fig:2}b).

Determinism ($DET$) is defined as the number of the recurrence points that form a  diagonal line of minimum length $l_{min}$ compared to all the recurrence points.
\begin{equation}
DET=\frac{\sum_{l=l_{\text{min}}}^{N}lP(l)}{\sum_{l=1}^{N}lP(l)},
\end{equation}
here $P(l)$ is the histogram of the line segments of a given length $l.$ Chaotic behaviour results in short diagonal line segments, nevertheless, periodic and quasi-periodic dynamics cause longer diagonals. Therefore, $DET$ can be used as a measure of the predictability of the system. Considerable care must be taken when $l_{\text{min}}$ has to be chosen too large, since then the histogram $P(l)$ can be very sparse and the parameter $DET$ becomes unreliable. A more complete list of RQA measures can be found in \cite{Marwan2007}.  
\subsection{R\'enyi (or correlation) entropy}
\label{sec_ent}

The RP texture and RQA clearly show the correlation with the underlying dynamics. However, as mentioned above the outcome of both methods depends on the reconstruction parameters $m$ and $\tau$ as well as on the threshold value $\epsilon.$ Although the technique of RQA is useful to describe the RPs' structure quantitatively, its measures are not invariants of the system. Several studies carried out that invariants in dynamical systems, such as Lyapunov exponent \citep{Rosenstein1993,Kantz1994,Choi1999}, correlation dimension \citep{Grassberger1983,Faure1998}, correlation entropy \citep{Beck1993}, can be acquired from the recurrence matrix.

Unfolding the dynamics properly from recurrence times requires that at least the series of return times contains the same amount of information generated by the complex system \cite{Baptista2010}. This information can be quantified by an entropy like measure. Besides the pioneering information-theoretic concept of entropy \citep{Shannon1949} a useful definition, fitted to dynamical systems, was developed by Kolmogorov \citep{Kolmogorov1965} and Sinai \citep{Sinai1959} (KSE). KSE, also called \textit{metric entropy,} in the ergodic theory of dynamical systems can be considered as the analogue of Shannon's block entropy differences, i.e. the entropy of the set per unit of time, in discrete-valued processes \citep{Lesne2014}. KSE is one of the most powerful invariant quantities to characterize chaotic dynamics \citep{Kantz2003}. For deterministic dynamics, \cite{Pesin1997} proved an inequality between KSE and the sum of positive Lyapunov exponent, $\mathrm{KSE}\geq \sum \lambda_i,$ where $i,\lambda_i\geq 0.$  Calculating the KSE for continuous systems with infinitely many states is a difficult task. The R\'enyi entropy (REN) (also called correlation entropy) is an invariant measure of dynamics and also the lower bound of KSE. Furthermore, REN can be calculated from return times without the  knowledge of a trajectory. This fact is ideal for our purposes.  

In brief, the R\'enyi entropy can be obtained from a recurrence matrix as \citep{Marwan2007} 
\begin{equation}
\mathrm{REN}(l,\epsilon)=-\frac{1}{l\Delta t}\ln P(l,\epsilon),
\label{eq:REN}
\end{equation} 
where $P(l,\epsilon)$ is the distribution of the diagonal lines of length at least $l,$ $\Delta t$ is the time between data points. Basically, $\mathrm{REN}$ measures the possible trajectories from a state up to time $l$ during the system is evolving. For example, a periodic orbit results in only one well-defined trajectory, therefore, $\mathrm{REN}=0.$ In other words, there is no uncertainty, a possible interpretation of entropy, in its future state. In contrast, for a sequence of independent and identically distributed random numbers, $\mathrm{REN}$ tends to infinity, since the number of possible trajectories due to the stochastic nature of dynamics increases very rapidly. Considering the limiting cases, one can state that quasi-periodic dynamics with low complexity yield $\mathrm{REN}\approx 0,$ while chaotic behaviour shows a positive but finite value of R\'enyi entropy. 

For completness, we note that the dynamical map (\ref{fig:1}) contains the initial conditions that lead to system disruption. That is, one of the members leaves the system and then the integration stops earlier than the default value. Since, for consistency, we want to compare the regular and chaotic time series with the same length, in this case the indicator megno is set to a large number (9999.0 -- white pixels in heat map) and also the entropy of corresponding RV time series is omitted from the analysis. Chaotic orbits, that are also unstable in some sense, surviving the integration time are confined to a finite region of the phase space. Indeed they often cover larger area than the ordered trajectories. Consequently, one can consider the phase space recurrences (although less frequently) of those trajectories, too, and use the entropy-like measure(s).

\section{Results}
\label{sec_results}

In this section, we present the REN as an indicator of stability obtained in the SJS system. We use certain line segments of the ($a,e$) parameter plane comparing REN and the structures presented in Fig.~\ref{fig:1}. The integration time is $10^3$ Jupiter periods (ca. 12000 years) with a sample of 25000 points, i.e. roughly 25 points per revolution (if not stated otherwise). The analysis and the construction of RPs and their statistics are restricted to one scalar time series, namely the x component of the Sun's barycentric velocity vector. This allows us to investigate the SJS configuration as it would be a remote exoplanetary system with a central star and two inhabitant giant planets.      

\begin{figure}
\begin{center}
\includegraphics[width=\columnwidth]{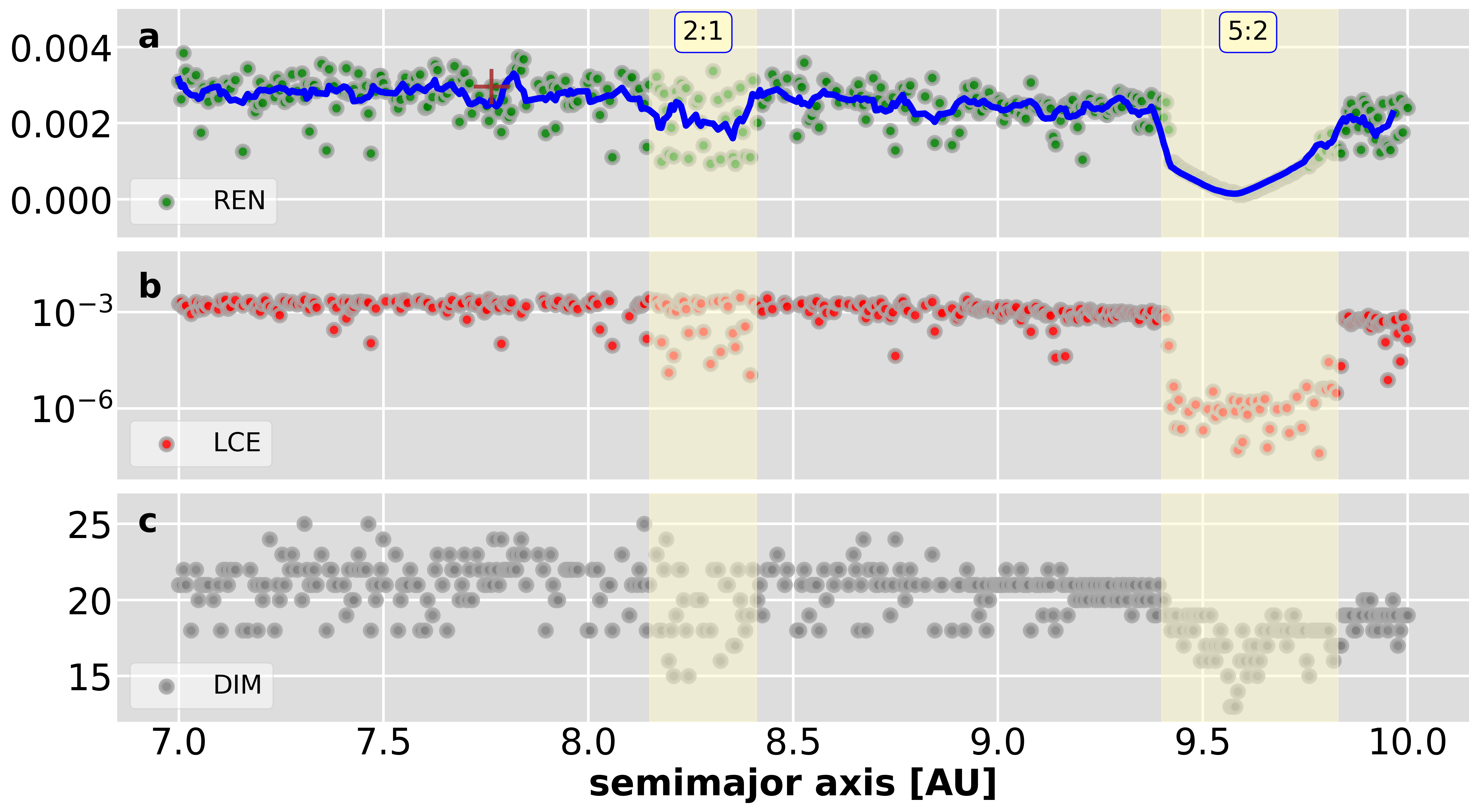}
\end{center}
\caption{(a) R\'enyi entropy along the horizontal line segment ($e=0.35$) in Fig.~\ref{fig:1}. Ordered motions taking place in resonances (yellow stripes) show smaller values than chaotic ones. Interestingly, the entropy measure captures the (fine) structure of the resonance. That is, its value has a minimum at the middle of the resonance indicating the less complexity of the periodic motion compared to quasi-periodic. The red cross at $a=7.75\mathrm{AU}$ is the point where the two dashed lines intersect in Fig.~\ref{fig:1}, also the initial condition of trajectory that is the base of RP in Figure~\ref{fig:2}. (b) Lyapunov characteristic exponents for the same initial conditions. (c) Embedding dimensions for different kinds of dynamics along the line segment used above. Unstable motion requires larger dimensional space to embed into.}\label{fig:3}
\end{figure}

Figure~\ref{fig:3} portrays two dynamical invariants, R\'enyi entropy (a), and Lyapunov characteristic exponent (LCE) (b) obtained from MEGNO, along the line segment of $e=0.35$ in Figure~\ref{fig:1}. The horizontal dashed line (Fig.~\ref{fig:1}) penetrates deep into the chaotic part in the ($a,e$) plane. It intersects the 5:2 mean motion commensurability at $9.4\lesssim a\lesssim 9.7.$ and also passes by the 2:1 MMC around $a\approx 8.2.$ 

The green points in panel (a) show the REN values of 500 initial conditions uniformly chosen from the interval $7\mathrm{AU}\leq a \leq 10\mathrm{AU}.$ The blue solid line illustrates the sliding window of size 7 to highlight the variation. Yellow stripes shade the position of MMCs to guide the readers' eyes. It turns out that REN, as its value comes close to zero, catches accurately the regular domain corresponding to 5:2 MMC. Moreover, it ''feels'' the remnant of the 2:1 MMC, too. Although, this salience is less prominent. It can be shown that when the reference line (horizontal in Fig.~\ref{fig:1}) is shifted down in eccentricity values, the indention at 2:1 MMC is getting deeper.  Figure~\ref{fig:3}b collects the LCEs for the same initial conditions as in panel (a). In a reassuring way, one can observe the similar texture in both pictures.  

In Section~\ref{sec_methods} we have seen the technique of time delay embedding that enables the reconstruction of the phase space trajectory from a single scalar time series. Although the embedding dimension does not carry any invariant properties about the underlying dynamics, the similarity between panels (a), (b) and (c) is clear. The idea behind the fact that embedding dimension copies the feature of dynamical quantities can be understood as follows. The more irregular the dynamics, the less the correlation between its elements. In other words, a motion with increased random elements (i.e. stronger instability) may require a larger embedding dimension than ordered motion \citep{Stergiou,Kovacs2019,Kovacs2020}.  

\begin{figure}
\begin{center}
\includegraphics[width=\columnwidth]{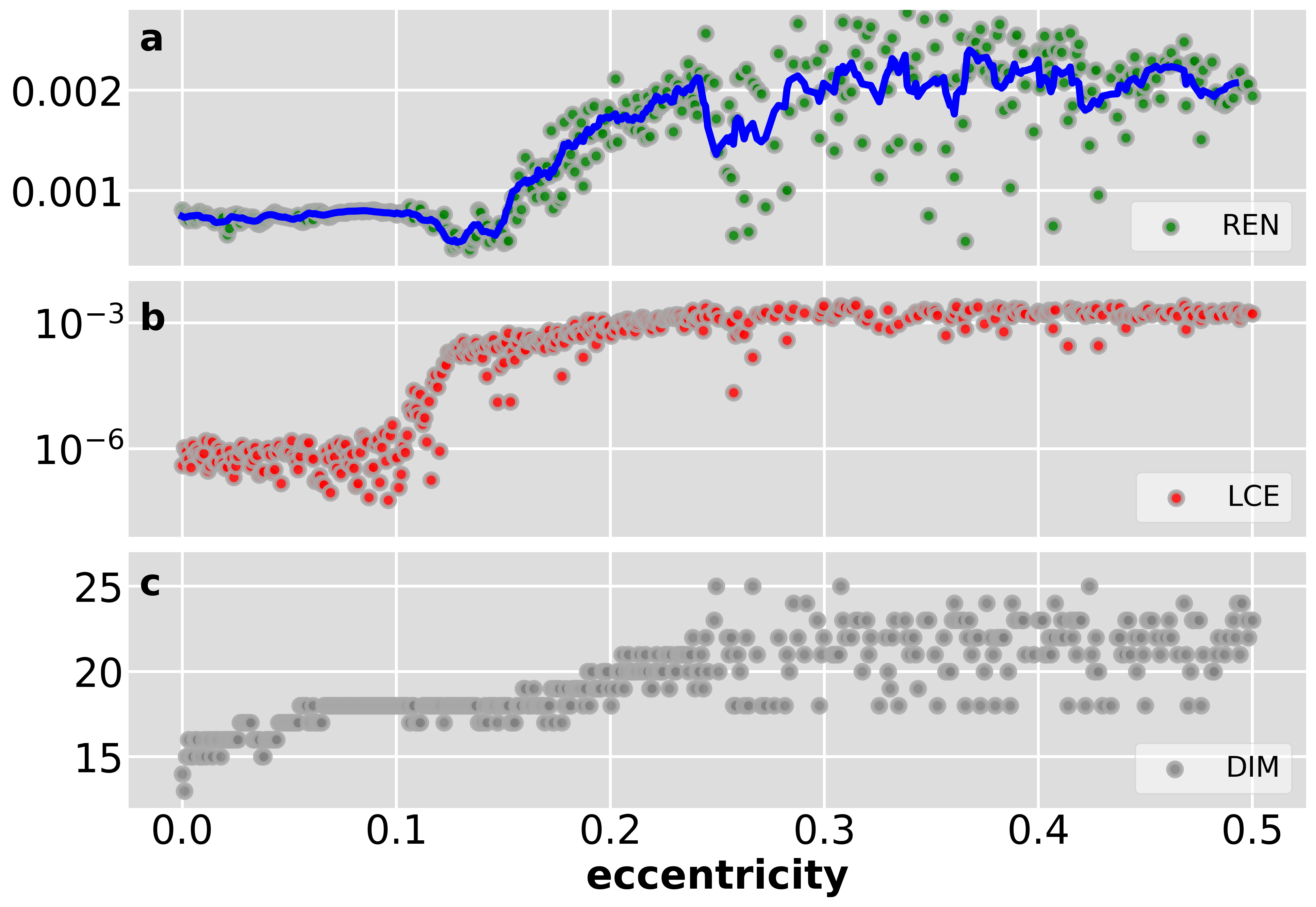}
\end{center}
\caption{The same measures as in Figure~\ref{fig:3} along the vertical line segment $a=7.75\mathrm{AU}.$ This calculation was made with RR=0.025.} \label{fig:4}
\end{figure}

For completeness, we give the REN along the line segment perpendicular to that one presented in Figure~\ref{fig:3}. In this case, the vertical dashed line ($a=7.75\mathrm{AU},$ Fig.~\ref{fig:1}) starts in the regular domain and, as eccentricity increases, it reaches the unstable part of the phase space. Choosing a lower RR we are able to ditinguish more robustly the ordered and chaotic domains. One can observe a slow cross-over from the ordered to the chaotic region for $0.12\lesssim e\lesssim 0.2$ which is also presented in panel (c) for embedding dimensions but missing for LCEs (panel (b)). Interestingly, the dip around $e\approx 0.12-0.15$ can be related to stable chaos or stickiness effect. Although there are many efforts to explore the stickiness via recurrences (see e.g. \cite{Zou2016}), we did not concentrate on that in present work. This phenomenon needs further investigations in the future.

\begin{figure}
\begin{center}
\includegraphics[width=\columnwidth]{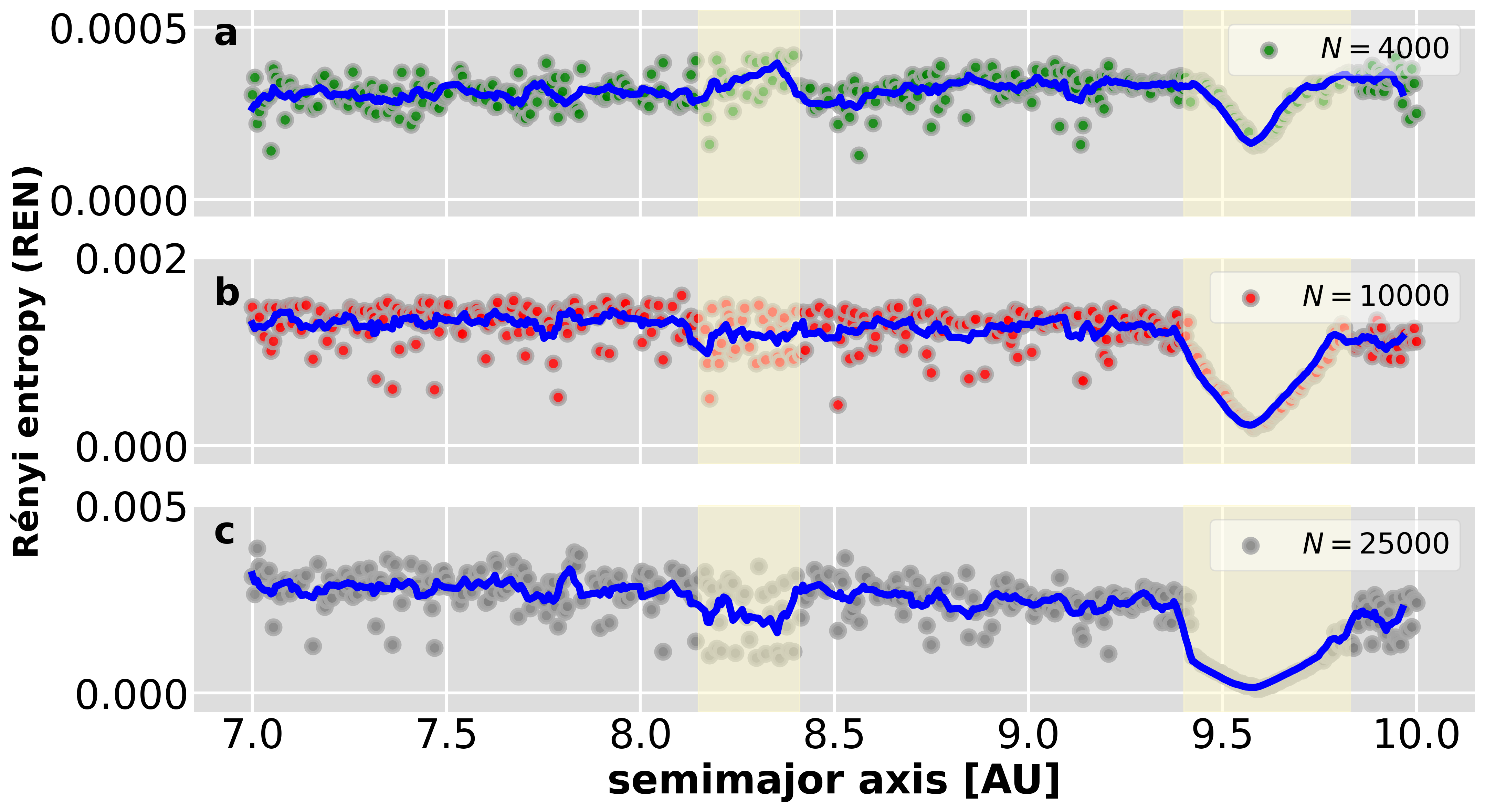}
\end{center}
\caption{The same integration time ($T=1000$ Jupiter revolutions) with different sampling rates. (a) N=4000 data points in radial velocity time series yielding 4 measurements per period. The contrast between the two types of motion in REN values is less since the information loss is due to the lower sampling frequency. (b)-(c) N=10000 and 25000 data points, the latter is identical with panel (a) in Fig.~\ref{fig:3}. }\label{fig:5}
\end{figure}

In the following, we are interested in the effect of sampling on REN. Therefore, we keep the integration time 1000 Jupiter periods and lower the frequency of measurements. Figure~\ref{fig:5} depicts the stability measure REN based on the sun's radial velocity including 4000 (a), 10000 (b), and 25000 (c) data points. As expected, we lose information about the dynamics as the sampling is getting poorer. Although, the influence of the resonance is still apparent. We, however, emphasize that 3-4 observations per rotation period are still optimistic in exoplanet spectroscopy, especially for hot Jupiters, Saturns (with a couple of days rotation period) where the signal is robust and therefore precise enough.      

\begin{figure}
\begin{center}
\includegraphics[width=\columnwidth]{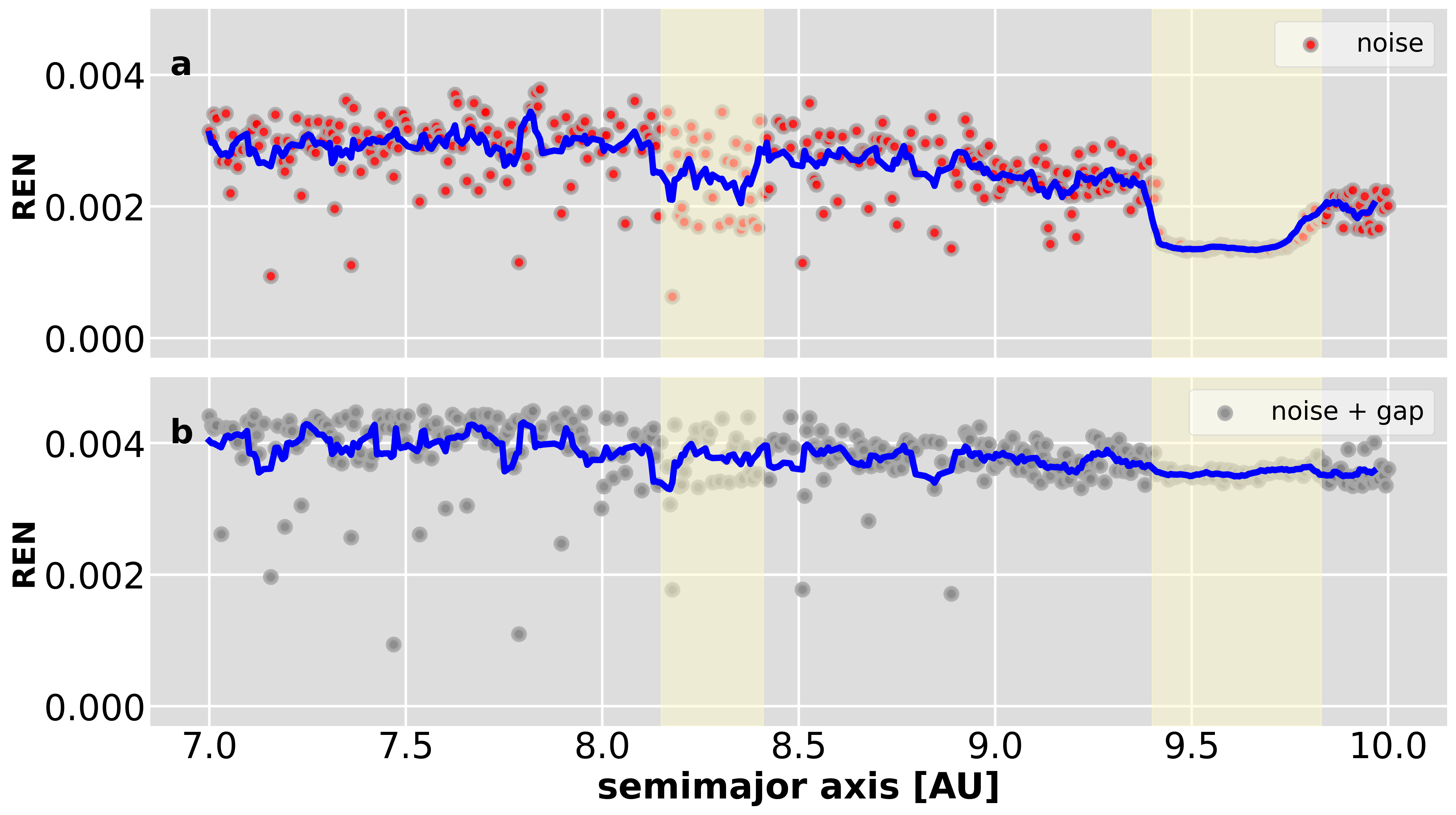}
\end{center}
\caption{The effect of noise (a) and missing data points (b) to the REN values. }\label{fig:6}
\end{figure}

Astronomical observations always contain measurement noise (e.g. atmospheric turbulence, CCD photon statistics, etc.) that cannot be omitted from the analysis. To mimic the consequence of such a disturbance to the stability indicator REN, we added 15\% Gaussian white noise to the original synthetic data set. Furthermore, random intervals, involving 20\% of data points in total, are also deleted from the time series as the result of seasonal observability or weather conditions. In Figure~\ref{fig:6} we present the REN values along the usual semimajor-axis range with additional observation noise (panel (a)) to the radial velocity data and randomly removed points from it (panel (b)). The entropy calculations seem to be robust against the noise since the decline at MMR is still visible although it is less sharp. Gaps in observations make the results more unreliable. Basically, there is no difference between the REN values of chaotic and regular motions.   

\section{Discussion}
\label{sec_discussion}

The above findings show that the second order R\'enyi entropy can be used as a powerful tool to analyse order and chaos in planetary systems from observation data. The advantage of this framework is two-fold. First, it does not require either the knowledge of the equations of motion or the exact number of planets in the system. The main assumption is that due to the coupling of variables, the dynamical information is encoded in state vector components and it can be extracted by an appropriate phase space reconstruction. Second, the analysis can be performed on relatively short and noisy time series which is essential in real-world observation. 

\begin{figure}
\begin{center}
\includegraphics[width=\columnwidth]{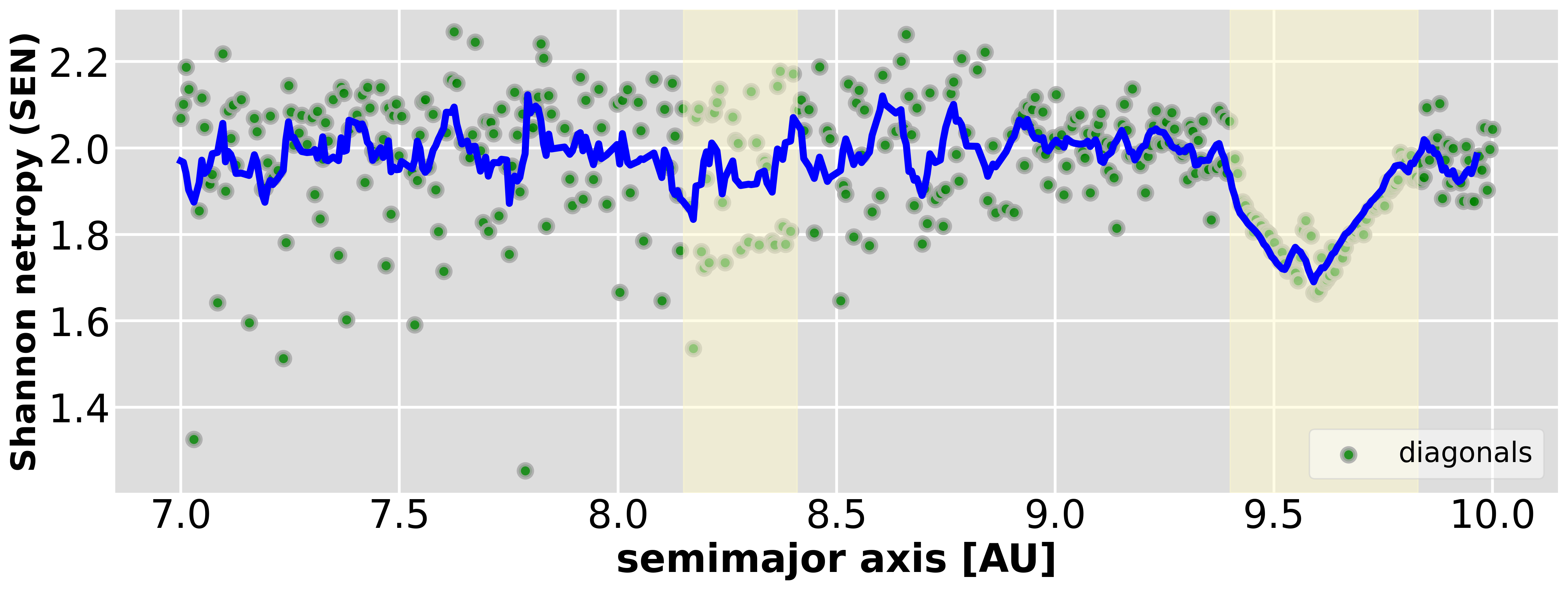}
\end{center}
\caption{Shannon entropy (SEN) calculated from the probability distribution of diagonal line segments of recurrence plots along the usual section ($7\mathrm{AU}\leq a\leq 10\mathrm{AU},\; e=0.35$) of the stability map. }\label{fig:7}
\end{figure}

Recently, promising attempts have been made by various groups \citep{Cincotta2018,Beauge2019,Cincotta2021,Kovari2022} to apply entropy measures to dynamical astronomy. At this point, we briefly discuss how to obtain RP-based Shannon entropy \citep{Marwan2007}. Shannon entropy of the diagonal line distribution reads
\begin{equation}
SEN=-\sum_{l=l_{min}}^{N}p(l) \ln p(l),
\end{equation}
where $p(l)=P(l)/N_l$ is the probability to find a diagonal line with length exactly $l$ in RP. Since the diagonal line segments can be considered as the occupation of the same part of the phase space for shorter or longer time scales. Its distribution is, therefore, compatible with the trajectory visit of partitioned the phase space, exactly how Shannon's entropy is defined. Figure~\ref{fig:7} indicates SEN for those initial conditions we have seen in previous sections. The essential structure is close to those obtained by REN, yet the contrast is somewhat smaller. Furthermore, we shall notice that recurrence plot-based REN is not independent of the threshold value $\epsilon.$ In recurrence quantification analysis, i.e. the statistical description of recurrecne plots (RPs), it is crucial to have ''the right amount of points'' in the recurrence plots. That is, having small recurrence rate (the number of dots compared to the empty spaces in RP, see Eq.~\ref{eq:rr}), we lose most of the information since the lack of texture in RP. In the opposite case (large recurrence rate), the RP is too crowded to gain any relevant information about the dynamics. The threshold parameter $\epsilon$ controls that the recurrence rate falls into the interval of 5-15\%. There are many studies about how to find the relevant $\epsilon,$ one of these says that the standard deviation of the time series provides a suitable tolerance parameter. Therefore, one might use it with care to describe the stability of motion.

\section*{Acknowledgements}
We acknowledge the computational resources and the fellowship (EK) for the Wigner Scientific Computing Laboratory (WSCLAB) of the Wigner Research Centre for Physics. EK is supported by the \'UNKP-21-3 New National Excellence Program of the Ministry for Innovation and Technology from the source of the National Research, Development and Innovation Fund. The authors thank the constructive discussions and suggestions to an anonymous referee, whose suggestions helped us to improve our manuscript considerably.

\section*{Data availability}
The data underlying this article will be shared on reasonable request to the corresponding author.




\bibliographystyle{mnras}
\bibliography{references} 







\bsp	
\label{lastpage}
\end{document}